# Polystyrene clusters captured by acoustic tweezers spontaneously rupturing

Saeedeh Kabiri[1], Amir Ghavidel[1], Sanaz Derikvandi[2], Fatemeh Rezaei [2], Ahmad Amjadi[1]

1. Department of Physics, Sharif University of Technology, Tehran 11155-9161, Tehran, Iran

2. Department of Physics, K. N. Toosi University of Technology, 15875-4416, Shariati, Tehran, Iran.

**Abstract**

Numerous investigations have demonstrated that standing acoustic waves can trap particles that range in size from microns to millimeters. Powerful tweezers may trap clusters of particles rather than single ones because their trapping radius is substantially larger than the size of the trapped particle. In this study, clusters of polystyrene particles measuring 450 microns in size that were suspended in an ionic surfactant solution were trapped at the nodes of acoustic standing waves. The correlation between surfactant concentration and threshold radius is examined, and potential mechanisms that could be responsible for the phenomenon are investigated. The findings demonstrated that adding polystyrene made clusters unstable and caused them to spontaneously rupture. Additionally, studies revealed that the cluster began to undergo spontaneous sequential ruptures after its radius above a particular threshold.

*Corresponding author: fatemehrezaei@kntu.ac.ir*

# Introduction

Acoustic tweezers are the ultrasonic devices which can trap micrometer to millimeter-sized particles using a high-gradient acoustic field. Since the first time that latex particles and frog eggs were trapped with two collimated focused ultrasonic transducers[1], numerous types of acoustic tweezers have been designed and fabricated. Some of the acoustic tweezers are constructed based on the standing waves. Furthermore, recent types of these tweezers employ more than two individually addressable piezoelectric elements to generate arbitrary wave fields and subsequently, permit the researcher to manipulate particles with more freedom[2-5]. Some others utilize highly focused single beams to trap and manipulate particle[6-8]. The first single beam acoustic tweezer was inspired by optical tweezers[1]. Another innovative technique of trapping with single beam acoustic tweezer is using a multi-foci Fresnel lens to create an acoustic Bessel beam[9]. Regardless of technical details, what is in common between all these tweezers is creating regions with high acoustic pressure gradient in the medium and trapping particles in these areas.

Among all the different types of acoustic tweezers, standing-wave tweezers have gained the most attention in the literatures [2-5,10-16]. This is due to the fact that standing acoustic fields produce strong pressure gradients which can trap particles more efficiently. More importantly, standing waves are very easy to be constructed. They can be generated either by twin transducers facing each other, or by a single transducer and a reflecting surface [17,18]. Moreover, the position of the particles can be easily controlled by adjusting the acoustic resonant frequency [12] or the relative phase of pressure waves [19,20]. Advent of interdigital transducers (IDTs) has even added to the popularity of standing-wave acoustic tweezers. IDTs generate surface acoustic waves (SAW) with the frequencies of tens to hundreds of megahertz. Therefore, they can manipulate very tiny particles, such as cells, with a high spatial resolution[12,21]. Different groups of researchers have reported trapping $10 \mu m$ particle using IDTs and moving them in x and y directions independently with high resolution [12,15]. It has also been shown that IDTs can manipulate particles as small as $1 \mu m$ in three dimensions [15]. Nowadays, due to the biocompatibility of acoustic waves, IDTs are widely used in lab-on-chip devices for cell manipulation and tissue-engineering [2-4].

It is mostly considered a privilege for the acoustic tweezers to able to trap single particles, although trapping clusters of particles rather than single ones has its own advantages. When particles approach each other in a high-gradient trap, the interparticle forces, which are negligible in long distances, become significant. It means that such powerful tweezers, in which the trapping size is more than the size of particles, are useful devices for studying the particle's interactions.

Different research groups have studied on interparticle forces in acoustic tweezers. Interparticle force was first introduced by Bjerknes in 1906 [22]. Bjerknes tried to calculate the primary and secondary radiation forces on the bubbles in a sound wave. His calculations on interparticle forces, and as well as most of the following researchers, were based on the idea of the rescattering of the acoustic waves from each particle [23-32]. They mentioned that the rescatterings occur infinitely, but the calculations are truncated at a finite number of times, mostly at the second order. In 1997 Doinikov et al [28] showed that when the viscous wavelength was comparable to the bubble radius, a new term would be added to the Bjerknes force, due to the acoustic streaming around the bubbles. In addition, in 2001, they [33] obtained an analytical expression for the time average of the acoustic radiation force among N particles floating in an ideal compressible fluid. In 2014, Silva and Bruus [30] completed the works of Doinikov et al. So that they found that in the Rayleigh range, where the size of particles is much smaller than the wavelength of sound, the acoustic interaction force among particles is well approximated by the gradient of the interaction potentials, and there is no limit on the interparticle distance. Finally, they managed to obtain the acoustic interaction force on a floating particle in standing or traveling plane waves. More interestingly, in 2015 Sepehrirahnama et al. [34] tried to solve the scattering problem for all the particles, simultaneously. Their methodology proposed multipole series expansion and weighted residual method to solve the governing Helmholtz equation employing the necessary boundary conditions on the particles' surfaces. It should be mentioned that their approach was more accurate compared to the previous researches, since it imposed the boundary conditions on all the particles simultaneously. In 2016, Sepehrirahnama et.al. [35] calculated interparticle radiation force between rigid spheres in a viscous fluid using multipole-Stokeslet method. They also studied the effect of streaming and found that acoustic streaming contributes significantly to the interaction among small spheres in an acoustic trap. Moreover, they mentioned that for every couple of particles trapped at a node, there would be a zero total force distance. If the centers of the two spheres fell in this distance, they would tend to move away from each other, and if they were placed out of this region, they would move towards each other.

In the present study, the clusters of 450 μm polystyrene particles were trapped at nodes of a standing wave and their behavior in the trap was investigated. It was observed for the first time that under some specific circumstances, clusters undergo spontaneous ruptures. Here, the repulsive forces which drive the particles away from each other are studied using finite element models and experimental methods such as zeta potential test.

**Theoretical background**

When two objects, for example two spheres, are placed in the acoustic field, the total acoustic force exerted on them has two parts, known as primary and secondary forces. The primary force is generated due to the incident sound wave which pushes the particles towards the pressure node or antinode based on their properties, but the secondary force is produced by the scattering of waves from the other objects [23-32].

**A: Primary radiation force:** First analysis of the acoustic radiation force for non-interacting incompressible particles dates back to the work in 1934 by King [36]. Then, in 1955 Yosioka and Kawasima [37] calculated the forces for non-interacting compressible particles. Their work was admirably summarized and generalized in 1962 by Gorkov.

According to Gorkov expression, the acoustic radiation force exerted on a compressible spherical particle with radius $a$ in an inviscid fluid is given by [38]:

$$F^{rad} = -\nabla U^{rad},  \quad (1)$$

$$U^{rad} = \frac{4\pi}{3} a^3 \left[ f_1 \frac{1}{2} K_0 \langle P_{in}^2 \rangle - f_2 \frac{3}{4} \rho_0 \langle v_{in}^2 \rangle \right], \quad (2)$$

$$f_1(\tilde{K}) = 1 - \tilde{K}, \tilde{K} = \frac{K_p}{K_0}, \quad (3)$$

$$f_2(\tilde{\rho}) = \frac{2(\tilde{\rho}-1)}{2\tilde{\rho}+1}, \tilde{\rho} = \frac{\rho_p}{\rho_0}, \quad (4)$$

where, $K_p$, and $K_0$ are bulk modulus of the particle and the fluid, respectively. In this equation, $\langle v_{in} \rangle$ and $\langle p_{in} \rangle$ indicate the time-average of the particle velocity and the pressure of the incident acoustic wave over one period time. For a standing wave, with the acoustic pressure, $p_{in}$ is expressed as:

$$p_{in}(z,t) = p_0 \cos(kz) \sin(\omega t), \quad (5)$$

and the potential of the acoustic velocity is calculated as below:

$$\varphi_{in}(z,t) = \frac{p_0}{\rho_0 \omega} \cos(kz) \cos(\omega t). \quad (6)$$

Moreover, the radiation potential is obtained as:

$$U^{rad} = \left[ \frac{f_1}{3} \cos^2(kz) - \frac{f_2}{2} \sin^2(kz) \right]. \quad (7)$$

and hence, the force exerted on the particle is calculated as follows:

$$F_z^{rad} = -\partial_z U^{rad} = 4\pi \Phi(\tilde{k}, \tilde{\rho}) k a^3 E_{ac} \sin(2kz), \quad (8)$$

$$E_{ac} = \frac{p_0^2}{4\rho_0 c_0^2}, \quad (9)$$

$$\Phi(\tilde{k},\tilde{\rho}) = \frac{1}{3}f_1(\tilde{k}) + \frac{1}{2}f_2(\tilde{\rho}) = \frac{1}{3}\left[\frac{5\tilde{\rho}-2}{2\tilde{\rho}+1} - \tilde{k}\right] \tag{10}$$

here, $E_{ac}$, and $\Phi(\tilde{k},\tilde{\rho})$ are the energy density and the acoustic contrast factor, respectively. If the contrast factor is positive for a particle, such as polystyrene, it will be attracted to the nodes of the standing wave and if it is negative, the particle will move to the antinodes. These equations provide sufficient accuracy for the most acoustophoresis calculations, because in most experiments, there is a significant distance between particles and the main force which drives the particles is the primary acoustic radiation force, but when the particles close to each other they start to have interactions. That is where interparticle forces become more pronounced.

**B: Secondary radiation force:**

In the Rayleigh range, the acoustic interaction force between two spherical particles in a standing plane wave is expressed as [30]:

$$\varphi_{ext}(z) = \frac{v_0}{k}\sin[k(z-h)]. \tag{11}$$

$$\mathbf{F}_{int}^{rad}(r) \approx -\frac{4\pi E_0 k^2 a_p^3 a_s^3 k_p k_s}{9 k_0^2 r^2}\sin^2(kh)\mathbf{e}_r. \tag{12}$$

In the above equation, $E_0 = \frac{1}{2}\rho_0 v_0^2$, $\rho_0$ is the fluid density, $k$ is the wave number, $a_p$ is the radius of the first particle, $a_s$ is the radius of the second particle, $k_p$ is the compressibility of the first particle, $k_s$ is the compressibility of the second particle, $r$ is the distance between the center of the particles, $k_0$ is the compressibility of the fluid, and $h$ is the distance between the center of the particle and the pressure node. It is demonstrated that the accumulation zone is located in the direction of wave propagation, while the particles may attract or repeal each other in the transverse direction [30].

For a system consisting of N particles, the total velocity potential is calculated as follows [34]:

$$\Phi = \Phi_{in} + \sum_{i=0}^{N} \Phi_{sc}^{(i)}, \tag{13}$$

here, $\phi_{in}$ is the wave velocity potential of the incident wave, and $\phi_{sc}^{(i)}$ is the velocity potential of the scattered wave from the $i^{th}$ particle.

The total force applied on a floating particle in an inviscid fluid due to scattered waves is expressed as below [35]:

$$F^{(i)} = -\left[\int_{S_i} \langle p_2^{(s)} \vec{I} \rangle \cdot \mathbf{n} \, da + \int_{S_i} \rho_0 \langle \mathbf{u}_1^{(s)} \cdot \mathbf{n} \rangle \cdot \mathbf{u}_1^{(s)} \, da\right], \tag{14}$$

where, $n$ is the unit vector perpendicular to the surface of the particle and its direction is outward, $\vec{I}$ is the identity tensor, and $p_2^{(s)}$ is the second-order pressure. The first term in equation (14) is created by the scattered waves from all other particles which can be written in terms of the first-order scattered variables. Furthermore, the scattered velocity in equation (14) is expressed as follows [35]:

$$\mathbf{u}_1^{(s)} = \mathbf{u}_1 - \mathbf{u}_1^{(i)}. \tag{15}$$

The variable $\mathbf{u}_1^{(i)}$ can be calculated in term of the scalar velocity potential as below:

$$\mathbf{u}_1^{(i)} = \nabla \Phi_i. \tag{16}$$

**Materials and methods**

**A: Experimental set-up of the acoustic tweezers**

In this research, a single 57KHz ring transducer was used for generating the acoustic waves in the medium. Sinusoidal electric waves were produced by a signal generator. The signals were amplified up to 400V by a power amplifier and then, implemented on the transducer. In addition, electric impedance matching was used to perfectly match the transducer with the electric system. It should be mentioned that the heavy backing of piezoelectric element from one side and the acoustic impedance matching layer on the other side of the transducer ensured that most of the acoustic power of the piezoelectric element was transferred into the medium of interest. The transducer was attached to the bottom of the container and the surface of water functioned as a reflector. Reflection coefficient of the water-air boundary for the acoustic waves moving in water medium is 99.9%, and consequently, it is reasonable to choose it as a reflecting surface. Moreover, the distance between the element and the surface should be an odd multiple of $\lambda/4$ to construct a standing wave. Here, it was chosen to be 70.9 mm ($3\lambda + \frac{\lambda}{4}$). In order to produce reliable results, all the experiments were repeated 10 times and a CCD camera was employed to monitor the process of trapping and record data.

**B: Synthesis of the solution of Sodium dodecyl sulfate (SDS)**

As soon as polystyrenes enter the medium of water, they start to aggregate. Normally, researchers add dilute surfactant solutions to water to prevent this problem. Generally, surfactants have a hydrophobic tail and a hydrophilic head. When they are adsorbed at the interfaces of particles and water, they reduce the surface tension between them and prevent aggregation. The hydrophilic head of each surfactant molecule is electrically charged which can be negative, positive, or neutral. Depending on the charge of the hydrophilic head, the surfactant is classified as anionic, nonionic, cationic or amphoteric. Anionic surfactants such as sodium dodecyl sulfate (SDS) are excellent for suspending polystyrene particles, which is why SDS is widely used in acoustic and optical tweezing of polystyrenes. As mentioned above, the hydrophilic head of SDS molecules is negatively charged. Hence, after adsorption on the surfaces of polystyrenes, they leave negative charge on them [39]. Changing the concentration of surfactant in the medium not only changes the amount of charge on the particles, but also varies the physical properties of the fluid such as viscosity and density. Host fluid plays a key role in acoustic tweezing experiments because all the vibrations induced by the source move through this medium and carry acoustic energy to the positions of particles. More importantly, total acoustic radiation force that each particle feels depends on the fluid and its physical properties, especially when the particles move very close to each other and have interactions.

In the current research, in order to study the effect of medium, different solutions of SDS (sodium dodecyl sulfate 70%) were constructed in deionized (DI) water with various concentrations. In all the stages of the experiment, all the devices and containers, which were in contact with the solution, were washed with DI water in advance, to prevent the entrance of any undesired ion into the medium. It should be noted that all the experiments were performed at room temperature.

The particles were 450 micrometer spherical polystyrenes. The density of polystyrene was just a bit greater than the density of water ($1050 \, kg/m^3$). So, they can be levitated easily in water.

**C: FEM Simulations**

In order to determine the magnitude and the direction of the primary and the secondary radiation forces applied on the particles, a simulation is performed to model the process of tweezing by solving the Helmholtz equation for the particles in the container. Finite element method (FEM) and MATLAB software were employed for this purpose. Once the Helmholtz equation is solved correctly, the acoustic pressure and the velocity are found in all the times and all the positions in the container.

The geometry of the model includes a three-dimensional tube with the length of $\frac{\lambda}{2}$ (13 mm) and two spheres with the radius of 450 μm. The sound propagation speed in this environment is about 1500 m/s, its density is 1000 $\frac{kg}{m^3}$, and the frequency is 57 kHz. The initial value of pressure is set to zero, because it is assumed that there is no pressure inside the tube before applying acoustic wave. In current research, it is intended to create an acoustic standing wave inside the tube. For this purpose, it is assumed that the plane wave enters from the wall located at $z = \frac{\lambda}{4}$ and it is considered that the wall is located at $z = -\frac{\lambda}{4}$ as a hard wall. Herewith, a pressure node is created in the center of the tube and pressure antinodes are produced on its walls. It should be mentioned that one of the particles is located at the center of the trap and another one has the distance of d with the first particle's center. Except for the wall through which the wave enters, Neumann's boundary condition is applied on all the boundaries. This means that the vertical derivative of the pressure changes on these boundaries is considered to be zero. This issue is also true for the surface of the particle, because the surfaces of the particles are assumed to be rigid.

In this paper, the investigated environment is discretized into 900,000 tetrahedral elements. The tetrahedral elements have a triangular cross-section which produces more reliable results in acoustics simulations. The maximum size of elements is set to be $4000 \mu m$ (i.e. $\frac{\lambda}{6}$). To make sure about the consistency of the obtained answers, the maximum mesh size on the particles' surfaces is chosen as $0.3 \mu m$. Moreover, on the cube surfaces, where less accuracy is needed, the maximum element size is $10 \mu m$.

In order to obtain the first-order pressure and velocity, it is needed to solve equation (4) to find the radiation force on the particles. For solving this integral, an integration surface must be defined first. Due to the constancy of the convective momentum, this surface can be the surface of the particle.

*Here, the simulations were performed using a Computer with RAM: 32GB, Processor Intel@CoreTM I7-4770k CPU@ 3.50GHz x 8, Graphics Intel@ Haswell Desktop, OS type- 64-bit, Disk 967.8 Gb done.

**Results and discussion**

By turning on the acoustic trap, polystyrene particles, suspended in the solution, migrate to the nodes of the standing wave. Due to the fact that the size of the trap zone is larger than the size of particles, clusters of the particles will be trapped at nodes, rather than single particle. Observations show that at very low concentrations of SDS (less than

0.5g/L) the clusters are quite stable at nodes, but at higher concentrations, when the radius of cluster gets bigger than a specific threshold, the cluster starts to undergo spontaneous sequential ruptures. Figure 1 presents the moment of the spontaneous rupture of a cluster. After each rupture, the particles get dispersed into the medium. The acoustic force of the trap pushes them toward the nodes immediately and makes them reform the clusters. In each rupture, some particles find the chance to escape from the trap and fall down due to gravity. This phenomenon is repeated several times until when the radius of the trapped cluster is smaller than a threshold and the cluster stops spontaneous ruptures. Moreover, experiments illustrate a significant change in the threshold radius, where the clusters start spontaneous rupture, with the increase of SDS concentration in the solution. Figure 2 represents the relationship between these two factors. Here, the experiments were repeated 10 times for each concentration and the average radius of clusters was measured by photography and the radius estimation codes in MATLAB.

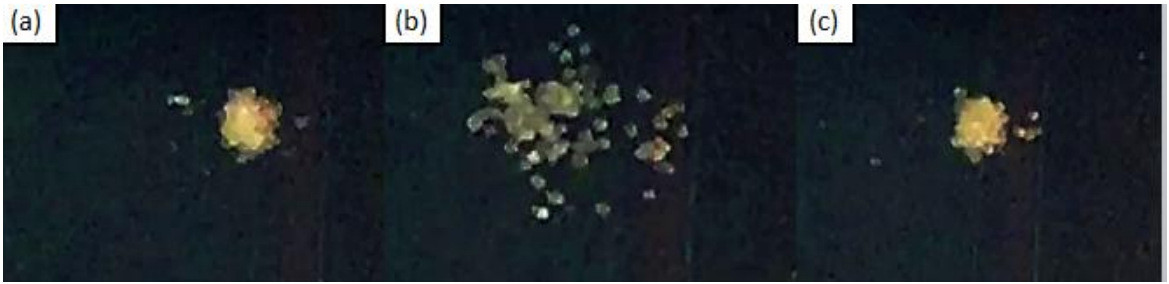

Figure 1. spontaneous rupture of polystyrene clusters at nodes. (a) Before rupture, (b) the moment of rupture, and (c) after rupture.

It should be noted that observations will not be explicable with a single theory. It seems to be a competition among different forces that induces the phenomenon. In the following sections, the various forces acting on the particles are investigated and the possible scenarios which can explain the phenomenon are discussed.

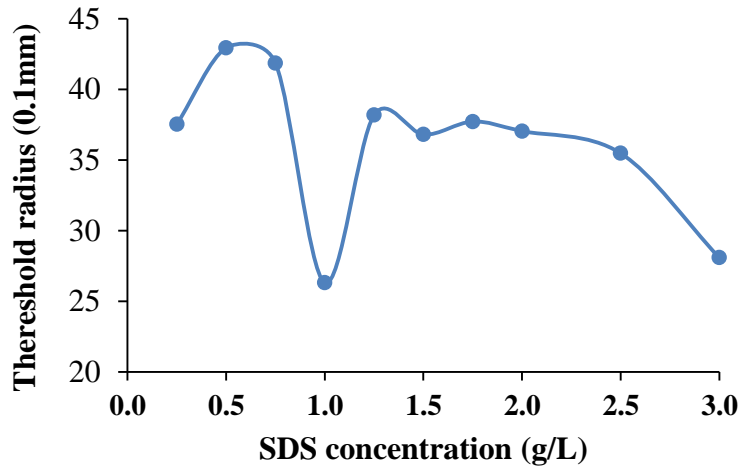

Figure 2. Threshold radius of polystyrene cluster versus. SDS concentration.

**Primary and secondary acoustic radiation forces**

Primary radiation force is the force exerted on the particles due to primary radiation of the acoustic source. When the particles are dispersed in the medium, they are far enough from each other and they do not feel interparticle forces caused by scattering of acoustic waves from other particles. In this situation, primary radiation force is dominant and the particles will be driven towards the nodes. When they aggregate at nodes and get very close to each other, interparticle forces become pronounced. In this paper, the acoustic interparticle force and the acoustic total force are investigated using a numerical method. Nevertheless, the experimental measurement techniques and the methods for theoretical calculations of the secondary force are difficult. Figure 3, represents the magnitude of the total radiation force acting on a pair of rigid spheres in an inviscid fluid in terms of the distance between two spheres. It can be clearly seen in figure 3 that the total radiation force graph consists of a discontinuity where the amount of the primary radiation force and the interparticle radiation force are equal and in opposite directions. Since this graph is depicted in the logarithmic scale, it can be inferred that the total radiation force is zero in discontinuities. In addition, each discontinuity represents a change in the direction of the total force, i.e., from repulsion to attraction. As long as the centers of the two spheres locate out of zero-force region, they experience an attraction towards the node and each other. As soon as the centers of the two spheres fall in the space between the pressure node and the place of the zero total force, both particles experience repulsion and tend to move away from each other. It is noteworthy that the repulsive force between the two spheres in very small distances is much more than the attraction they feel in large distances.

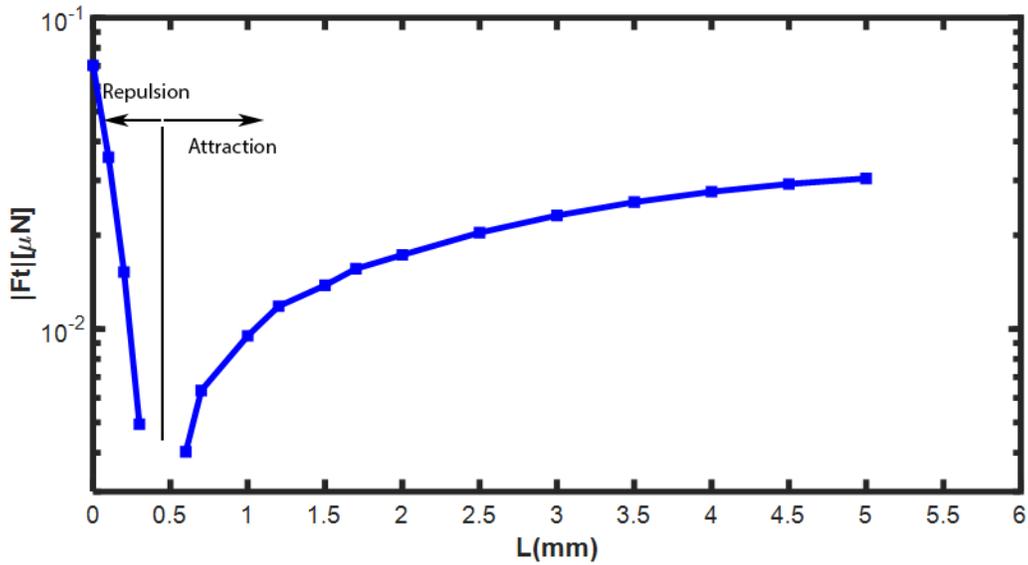

Figure 3. Total acoustic radiation force acting on two particles in the trap as a function of their distance.

Figure 4 demonstrates the distribution of the interparticle radiation force with distance. As the figure shows, if the particles are pushed towards each other from $L = 3.5mm$ to $L = 0.5mm$, the repulsive force between them will be increased by three orders of magnitudes, which is a considerable amount.

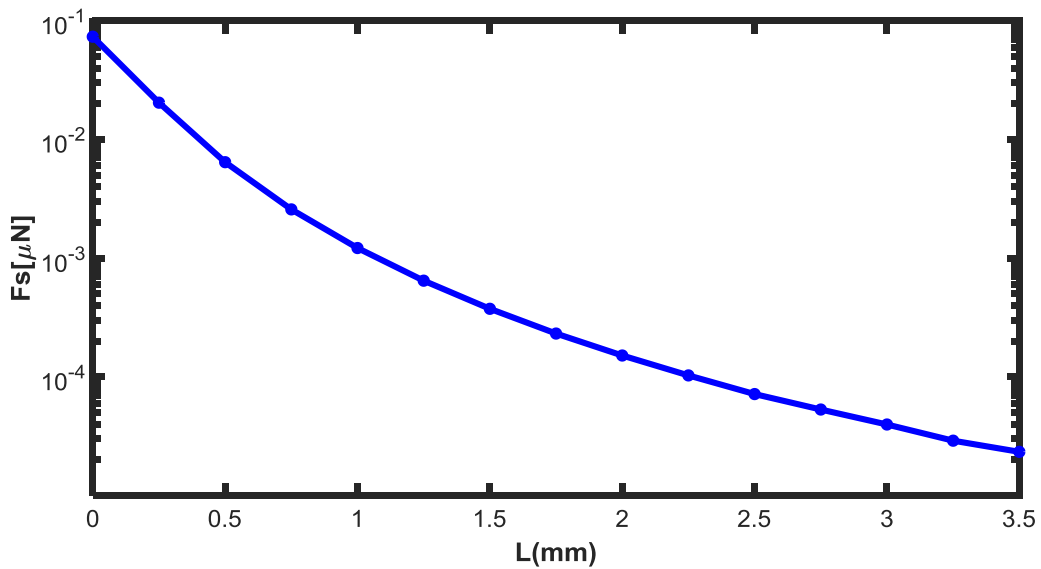

Figure 4. The secondary radiation force acting on two particles in the trap versus distance.

In a big cluster, particles of the outer layer, which are still away from the zero-total radiation force, tend to move towards the node and push the particles in the inner layer of the cluster inwards. It happens when the particles of the inner layer have got very close to each other and tend to move away from the node. Finally, the repulsion force between the particles of the inner layer dominates the attractive force which drives other particles toward the node and the cluster explodes.

As the concentration of SDS increases from 0.5 g/L to 3g/L, the viscosity of the fluid enhances [35]. Enhancement of the viscosity of the medium leads to the increase of interparticle force. This observation is in very good accordance with the results reported in figure 2. So, the repulsive force among the particles of the inner layer can dominate the pressure of the outer particles more easily and the cluster will get dispersed in the smaller radii. That is the reason why there is a general downward trend in the graph of figure 2.

Furthermore, as figure 2 illustrated, there is a minimum of threshold radius at the concentration of 1g/L, which means the clusters get very unstable at this specific concentration. In the following part, an investigation will be performed to find what happens at this point and discuss the reason of this unexpected instability.

**Electrostatic forces**

As it was mentioned before, the surfactant used in this experiment, SDS, is a kind of anionic surfactant. In anionic surfactants, the heads of surfactant molecules carry a net negative charge. Therefore, when they're absorbed on the surface of polystyrenes, transmit a net negative charge to the particles. Here, to evaluate the amount of the negative charge on the particles, zeta potential tests were performed on the polystyrenes immersed in different solutions of SDS with various concentrations. It should be noted that zeta potential is the potential difference between the dispersion medium and the stationary layer of fluid attached to the particle which is widely employed for quantification of the magnitude of the charge. Figure 5 shows the distribution of the zeta potential versus SDS concentration in the fluid.

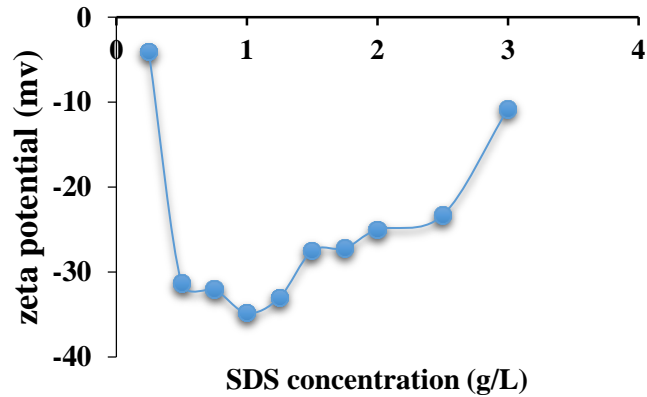

Figure 5. Zeta potential versus SDS concentration in the fluid.

As the figure 5 shows, at the concentration of 1g/L, the particles carry the utmost value of the negative charge. So, at this specific point, where the electrostatic repulsion is at its maximum, electrostatic repulsion intensifies the effect of acoustic repulsive force and helps the cluster break-up happen more quickly. It must be noted that the electrostatic repulsion can't be the only factor in forming cluster ruptures. This can be attributed to the fact that in figure 5 by increasing the concentration of SDS from 1g/L to 3g/L, the net amount of negative charge on the particle decreases. Hence, it is expected that the clusters become more stable in the fluids at higher SDS concentrations, but it is completely in contradiction with what was inferred from figure 2.

In addition, the data of figure 5 is in good accordance with what is expected from the chemistry of SDS solutions. As the concentration of SDS increases in the medium, more negatively charged molecules are adsorbed on the surface of the particles, but this adsorption stops at a specific concentration, known as Critical Micelle Concentration (CMC). When the surface of the particles is fully overlaid with SDS molecules, they start to form micelles, rather than trying to attach to the particles. So, after CMC point, not only increasing the concentration of SDS doesn't lead to the enhancement of negative charge on the particles, but also it helps the SDS molecules to form more micelles. This phenomenon contributes to the decrease of the net negative charge on the polystyrenes. Moreover, it is predicted by the observations that the CMC of SDS solution used in the experiments is just a bit more than 1g/L. According to the literature Ref [40], at room temperature, CMC of SDS solution in water is approximately about 0.2% of mass fraction or 0.008 mol/L. It means that the CMC of the solution used in the current research is about 2g/L, which is very close to what was expected from the experiment.

**Validation of the simulation results**

To investigate the contributions of the primary and secondary radiation forces in forming the spontaneous ruptures of clusters, it is needed to estimate the value of these forces in the experiment. In current research, finite element method(FEM) is employed for this purpose. To validate the accuracy of the suggested model, the results of the finite element model are compared with previous literatures [35]. In 2016, Sepehrirahnama et al. [35] proposed a numerical algorithm based on the multipole series expansion and Stokeslet method to calculate the secondary and the total radiation forces acting on a pair of spheres in the viscous and non-viscous fluids. To check the accuracy of the present model, a similar simulation is performed with the same dimensions and boundary conditions as Sepehrirahnama et al utilized. The only difference between these two models is that they used Stokeslet method by considering the viscous and streaming effects, while here, integration based on the finite element method is utilized for calculating the acoustic interparticle force in an inviscid fluid. It should be noted that in both models, two identical rigid spheres are located along the wave direction (axisymmetric configuration), and equally spaced from the pressure node. Moreover, the global coordinate system is placed at the pressure node. The standing wave is in the $z$ direction. The surface-to-surface distance between the two spheres is denoted by $L$ and the radius of each sphere is equal to 10μm. Due to the symmetrical configuration, the radiation forces acting on the spheres are equal and apply in opposite directions. For numerical calculations, the inviscid water in STP condition (standard temperature and pressure) is considered as the host fluid. The frequency of the standing wave is 1.5MHz and its wavelength is 1 mm. The pressure amplitude is set at 1 bar. Figure 6 presents a comparison between the data of the present model with Sepehrirahnama's reports. As it can be clearly seen in this figure, the data of the two models show approximately good agreement with each other, because the amount of the acoustic interaction force in an inviscid fluid has the same trend as in the viscous fluid. In addition, as it was expected, the magnitude of acoustic interaction force in a non-viscous fluid is less than the viscous fluid in which streaming effects have been taken into account. The difference between the results of the current model and the reports of Sepehrirahnama et al. can be attributed to the differences in methods of calculation.

Figure 7 shows a comparison between the total acoustic radiation force obtained by the present model and the reports of Sepehrirahnama et al [35]. As the figures illustrates, there is a good agreement between the results of the two models for ideal fluid. It is noteworthy that when viscosity of the fluid is considered, the repulsive force between the particles is dramatically enhanced.

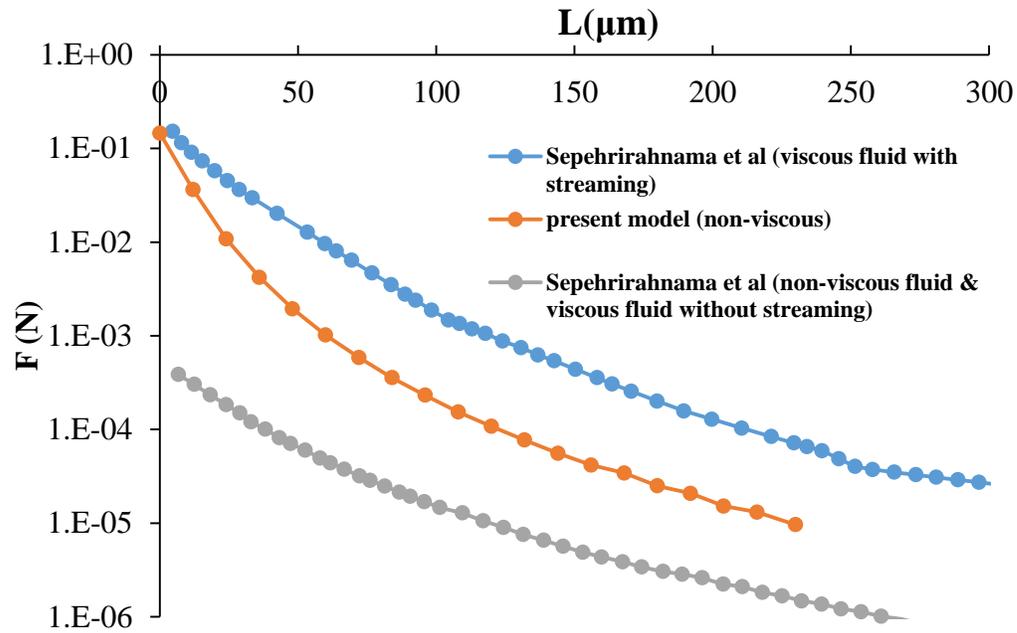

Figure 6. A comparison between the interparticles force obtained by the present model, and the reports of Sepehrirahnama et al [35].

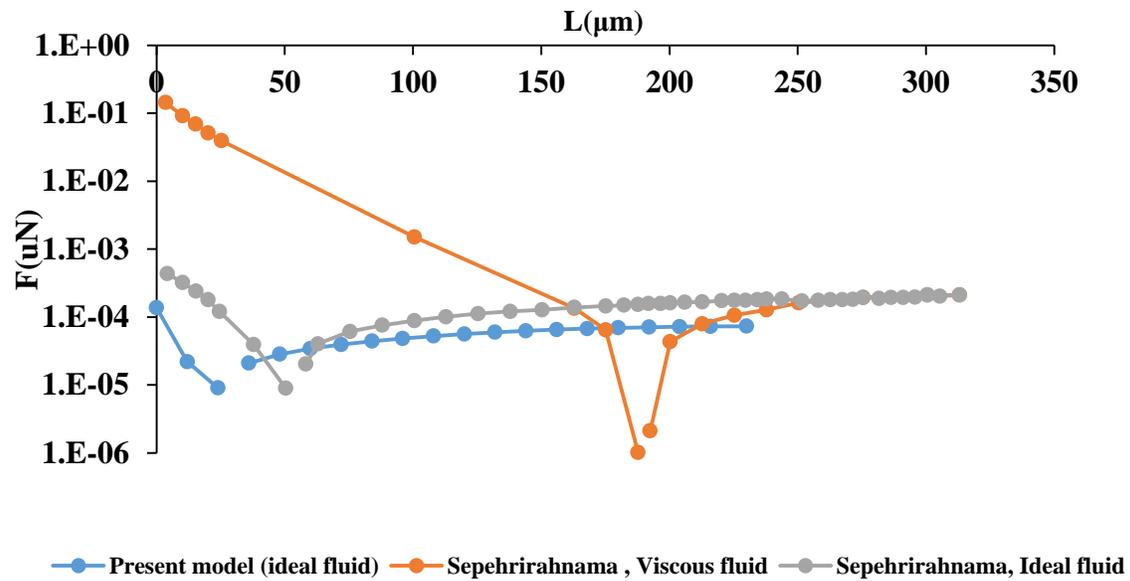

Figure 7. A comparison between the total acoustic radiation force obtained by the present model, and the reports of Sepehrirahnama et al [35].

## Conclusion

In this research, standing acoustic waves were employed to trap clusters of polystyrene particles and the interaction of particles in the trap was investigated. Dilute solutions of SDS with various concentrations were added to the medium of trapping and it was observed that after the concentration of 0.5 g/L, the clusters undergo sequential spontaneous ruptures. As the concentration of SDS increased, the threshold radius where the cluster starts exploding decreased. It was shown by finite element methods that the main reason of this phenomenon is the secondary radiation force, which results from the scattering of acoustic waves from other particles in the medium. FEM simulations predicted that for every couple of particles in the acoustic trap there is a specific distance where the direction of total radiation force suddenly changes from attraction to repulsion. When two particles approach the trapping point, at first, primary radiation force attracts them towards the node, but when they move closer, interparticle force dominates all other forces and drives them apart. Besides, it was observed in the experiments that at the concentration of 1g/L, the clusters become extraordinarily unstable. The threshold radius of the cluster at this point was found to be much less than all other points. It was shown by zeta potential tests that at this specific point the amount of negative charge on polystyrenes are maximum and it's the resulting coulomb force drives the particles apart at this concentration. Furthermore, the results of zeta potential tests verify that the coulomb force cannot be the main reason of spontaneous explosions of clusters.

**Author contributions:**

Saeedeh Kabiri and Amir Ghavidel designed and performed the experiments and prepared initial version of the manuscript with data analysis. Sanaz Derikvandi contributed to the simulation calculation. Fatemeh Rezaei supervised the manuscript and contributed on manuscript structure and data interpretation. Ahmad Amjadi managed the experimental section by checking the results. Saeedeh Kabiri and Sanaz Derikvandi wrote and reviewed the paper.

**Competing interests**

The authors declare no competing interests.

**Data availability statements**

The datasets generated or analyzed during the current study are available and presented just in this paper.